\documentclass[twocolumn]{aastex631}

\newcommand{\um}{{\rm \mu m}}
\newcommand{\E}[1]{\times 10^{#1}}

\newcommand{\MPopIII}{M_{\rm Pop III}}
\newcommand{\Esn}{E_{{\rm SN}}}

\newcommand{\abn}[1]{n_{\rm {#1}}}
\newcommand{\abM}[1]{M_{\rm {#1}}}

\newcommand{\abH}[1]{{\rm [{#1}/H]}}

\newcommand{\abFe}[1]{{\rm [{#1}/Fe]}}

\newcommand{\MetSilicon}{{\rm Si}}

\newcommand{\Enstatite}{{\rm MgSiO_3}}
\newcommand{\Forsterite}{{\rm Mg_2SiO_4}}
\newcommand{\Magnetite}{{\rm Fe_3O_4}}
\newcommand{\Silica}{{\rm SiO_2}}

\newcommand{\Alumina}{{\rm Al_2O_3}}

\newcommand{\Zsun}{Z_{\odot}}
\newcommand{\Msun}{M_{\odot}}

\newcommand{\Mcut}{M_{\rm cut}}

\newcommand{\Mdust}{M_{\rm d}}
\newcommand{\Mmix}{M_{\rm mix}}
\newcommand{\Mrem}{M_{\rm rem}}
\newcommand{\fej}{f_{\rm ej}}

\newcommand{\Mpr}{M_{{\rm pr}}}

\newcommand{\Mr}{M_{\rm r}}

\defcitealias{Nozawa03}{N03}
\defcitealias{Nozawa13}{N13}
\defcitealias{Chiaki15}{C15}
\defcitealias{Chon16}{C16}
\defcitealias{Chiaki17}{C17}
\defcitealias{Chiaki18}{C18}
\defcitealias{Chiaki19}{Paper I}
\defcitealias{Smith15}{S15}
\defcitealias{Hirano14}{H14}
\defcitealias{Ishigaki14}{I14}
\defcitealias{Marassi14}{M14}
\defcitealias{Marassi15}{M15}

\usepackage{xcolor}

\begin{document}

\title{Supernova dust synthesis --- Carbonaceous dust in the very early universe}

\author[0000-0001-6246-2866]{Gen Chiaki}
\affiliation{Department of Social Design Engineering, National Institute of Technology (KOSEN), \\
Kochi College, 200-1 Monobe Otsu, Nankoku, Kochi, 783-8508, JAPAN}
\affiliation{Division of Science, National Astronomical Observatory of Japan, \\
2-21-1 Osawa, Mitaka, Tokyo 181-8588, Japan}



\author[0000-0002-4343-0487]{Chiaki Kobayashi}
\affiliation{Centre for Astrophysics Research, Department of Physics, Astronomy and Mathematics, \\
University of Hertfordshire, Hatfield, AL10 9AB UK}

\author[0000-0001-8537-3153]{Nozomu Tominaga}
\affiliation{Division of Science, National Astronomical Observatory of Japan, \\
2-21-1 Osawa, Mitaka, Tokyo 181-8588, Japan}
\affiliation{Department of Astronomical Science, SOKENDAI (The Graduate University for Advanced Studies), \\
 Osawa 2-21-1, Mitaka, Tokyo, 181-8588, Japan}
\affiliation{Department of Physics, Faculty of Science and Engineering, Konan University, \\
8-9-1 Okamoto, Kobe, Hyogo 658-8501, Japan}



\begin{abstract}
Recent observations have revealed the spectral feature of carbonaceous grains even in a very distant galaxy.
We develop a state-of-the-art dust synthesis code by self-consistently solving molecule and dust formation
in supernova (SN) ejecta that contain various elements in different layers. 
With a progenitor mass $25~\Msun$ and explosion energy $10^{52}$ erg,
we run the following four test calculations to investigate the impact of input physics.
(i) With molecule formation solved, our SN model produces $8.45\E{-2}~\Msun$ carbonaceous grains.
(ii) If all available C and Si were initially depleted into CO and
SiO molecules, respectively,
the C grain mass could be underestimated by $\sim 40$\%.
In these two models producing $0.07~\Msun$ $^{56}$Ni without mixing fallback,
a large amount of silicates ($0.260~\Msun$) created in O-rich layers are also ejected and likely to hide the spectral feature of carbonaceous grains.
We then consider mixing-fallback
that can reproduce the observed elemental abundance ratios of C-normal and C-enhanced extremely metal-poor stars in the Milky Way. 
(iii) In the former, the mass ratio of carbonaceous to silicate grains is still small ($\sim 0.3$).
However, (iv) in the latter (known as a ``faint SN''), while the C grain mass is unchanged ($6.78\E{-2}~\Msun$), the silicate mass is reduced ($9.98\E{-3}~\Msun$). 
Therefore,
we conclude
that faint SNe can be a significant carbonaceous dust factory in the early Universe. 
\end{abstract}


\section{Introduction} \label{sec:intro}
Long-term observational campaigns with the Herschel Space Observatory and the Atacama Large Millimeter Array (ALMA) have revealed 
that high-redshift galaxies ($z\sim 8$) contain dust grains -- the condensates of metallic nuclei -- 
with a mass of $\sim 10^8~\Msun$, corresponding to the dust to stellar mass ratio of $\sim1$\%.
It is puzzling how these galaxies have acquired such a large amount of grains for the short time, $\sim 600$ Myr, since the Big Bang \citep{Watson15, Tamura19, dayal22, Witstok22}.
In addition, the James Webb Space Telescope (JWST) have opened a new window to observe the galactic chemical and dust evolution (GCE) in the very early Universe \citep[e.g.,][]{Bunker23,Witstok23b,DEugenio23,Carniani24}.

With the NIRSpec instrument on the JWST, \citet{Witstok23a} found the spectroscopic feature of carbonaceous (C) grains known as a ``UV bump'' in the galaxy, JADES-GS-z6-0, at redshift $z=6.7$. 
%
%
%
In such a distant galaxy, the origin of C grains is unknown; since there is no time for low-mass stars to contribute to the dust formation, core-collapse supernovae (SNe) from the first generation of metal-free (Population III or Pop III) stars are likely to be the major dust source \citep{Todini01, Schneider03, Nozawa03}.
However,
it has been considered that SNe mainly produce another species, silicates, because SN ejecta is oxygen-rich in most cases \citep{Bianchi07, Sarangi15, Marassi19}.
Unless the C grain mass is comparable to or more than the silicate grain mass, the remarkable UV bump cannot appear \citep{Nozawa13a}.
More sophisticated dust formation model is required to predict the mass of each grain species in detail.


In this paper, based on our previous work \citep[][\citetalias{Nozawa13}]{Nozawa13}, we revisit dust formation in Pop III SNe,
developing a state-of-the-art dust synthesis code with all relevant physical processes: 
hydrodynamics, explosive nucleosynthesis, radiative transfer, nuclear decays, self-consistent molecule formation, and ejection with mixing/fallback.
First, 
our code starts with pre-SN stars calculated by \citet{Umeda00} with their stellar evolution code.
Then, we calculate explosive nucleosynthesis \citep{tominaga07}.
It is important to note that these have also been used to calculate our nucleosynthesis yields for the Galactic chemical evolution in \citet{kobayashi06,kobayashi20} as well as for the Galactic archaeology studies in \citet{tominaga14} and \citet{Ishigaki18}.
These works found a significant contribution of supernovae with explosion energies of $\ge10^{52}$ erg, hypernovae.

The pre-SN stars have an ``onion-like structure'', where lighter elements lie in outer layers.
Hence, C grains and silicates are expected to mainly form in outer and intermediate layers of ejecta, respectively.
If the chemical composition of SN ejecta is fully mixed, a significant amount of C atoms would be oxidized into CO molecules, which can suppress C grain formation as pointed out by \citet{Marassi14}.
Therefore, we calculate dust formation keeping the radial structure of SN ejecta. 

Finally, we consider molecule formation that can compete with dust formation \citep{Marassi14, Marassi15}.\footnote{In this work, we define ``molecules'' as diatomic species that are not locked up into grains.}
For example, in the inner part of the C layer, oxygen is also synthesized. 
If C atoms are completely depleted into CO molecules, there are no longer C atoms available for forming C grains.
Similarly, metallic silicon (Si) grains form in Si-rich layers, where sulfur is also abundant, and thus SiS molecules can suppress Si grain formation.
\citet{Marassi15} did not consider SiS molecule formation.
\citetalias{Nozawa13} considered both, but, in their model, all C and Si atoms were assumed to be initially depleted into CO and SiO molecules, respectively, for simplicity.
This setup might lead to 
underestimate of 
C grain mass 
and overestimate of Si grain mass. 
To overcome this problem,
we here calculate huge chemical networks of not only grain formation but also molecule formation (Section \ref{sec:methods}).

Most of earlier works considered a SN model with a ``mass cut'' that yields a constant $^{56}$Ni mass of $\abM{^{56}Ni} = 0.07~\Msun$
\citep[\citetalias{Nozawa13},][]{ Biscaro14}.
This value was chosen to reproduce the observed light curve of SN1987A and is not necessarily applicable to Pop III SNe.
Although Pop III SNe have not been directly observed yet, 
their properties have been constrained from the observed elemental abundances of extremely metal-poor (EMP) stars, defined with their iron abundances $\abH{Fe} < -3$ \citep{Ishigaki18, Choplin19};\footnote{${\rm [A/B]} = \log (\abn{A}/\abn{B}) - \log (\abn{A}/\abn{B})_{\odot}$, where $\abn{X}$ is the number abundance of a element X. Here we use the solar elemental abundance presented by \citet{Asplund09}.}
EMP stars are considered to form in clouds enriched by Pop III SNe \citep{Chiaki18, Chiaki19, Chiaki20}.
In order to reproduce the elemental abundances of EMP stars, mixing and fallback is introduced in SN nucleosynthesis 
\citep{Umeda03, tominaga14}.
These processes can significantly suppress the production of grains made of heavier elements than O.

\begin{figure*} 
\epsscale{1.17}
\plotone{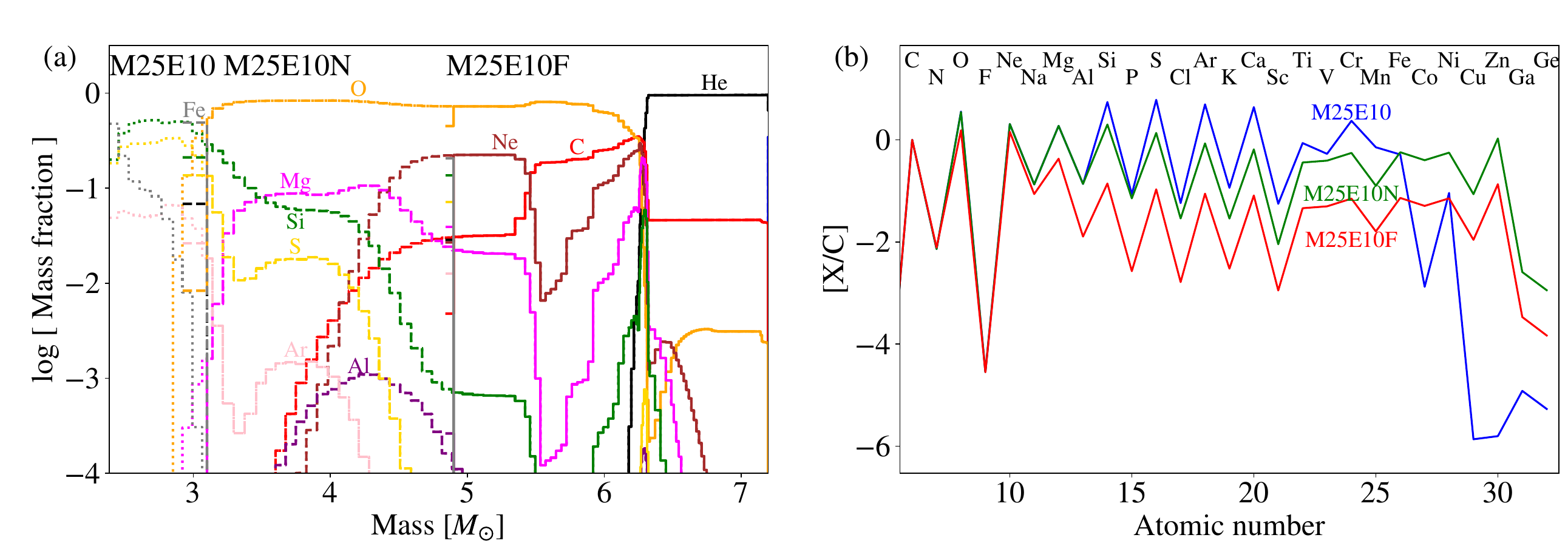}
\caption{(Panel a) Mass fraction of elements as a function of enclosed mass in the He core (outside the remnant) for a Pop III hypernova with a mass of 
$25~\Msun$ and explosion energy $10$ B.
The dotted, dashed, and solid curves show the results of the model without mixing/fallback ({\tt M25E10}), and 
the normal ({\tt M25E10N}) and faint ({\tt M25E10F}) hypernova models, which respectively reproduce the elemental abundances of C-normal and C-enhanced metal-poor stars.
(Panel b) Number fraction of elements from B to Ge relative to C in the models {\tt M25E10} (blue), {\tt M25E10N} (green), and {\tt M25E10F} (red).}
\label{fig:rx_ini}
\end{figure*}

\begin{deluxetable*}{ccccccccccccc}
\tabletypesize{\scriptsize}
\tablewidth{0pt} 
\tablecaption{Input parameters and dust masses of Pop III SN models\label{tab:models}}
\tablehead{
\colhead{Model} & \colhead{$\Mpr$}& \colhead{$\Esn$} & \colhead{$M_{\rm cut}$} & \colhead{$M_{\rm mix}$} & 
\colhead{$\log f_{\rm ej}$} & \colhead{$M_{\rm rem}$} & \colhead{$\abM{^{56}Ni}$} & \colhead{$t_{\rm fin}$} &
\colhead{$\abM{C}$} & \colhead{$\abM{\MetSilicon}$} & \colhead{$\abM{\Forsterite}$} & \colhead{$\abM{d}$} \\
\colhead{} & \colhead{($\Msun$)} & \colhead{(B)} & \colhead{($\Msun$)} & \colhead{($\Msun$)} & \colhead{} & 
\colhead{($\Msun$)} & \colhead{($\Msun$)} & \colhead{$d$} &  \colhead{($\Msun$)} & \colhead{($\Msun$)} & \colhead{($\Msun$)} & \colhead{($\Msun$)}
} 
\colnumbers
\startdata 
{\tt    M25E10}     &   $25$ &   $10$ &    $1.69$ &    $2.39$ & $-\infty$ &    $2.39$ &  $7.00\E{-2}$ & $ 840$ &  $8.45\E{-2}$ &       $0.196$ &       $0.260$ &       $0.926$  \\
{\tt M25E10mol.fix} &   $25$ &   $10$ &    $1.69$ &    $2.39$ & $-\infty$ &    $2.39$ &  $7.00\E{-2}$ & $ 840$ &  $5.77\E{-2}$ &       $0.270$ &       $0.260$ &       $0.985$  \\
{\tt   M25E10N}     &   $25$ &   $10$ &    $1.69$ &    $3.1$  &   $-0.9$  &    $2.92$ &  $8.41\E{-2}$ & $ 713$ &  $8.70\E{-2}$ &  $4.43\E{-2}$ &       $0.259$ &       $0.706$  \\
{\tt   M25E10F}     &   $25$ &   $10$ &    $1.69$ &    $4.9$  &   $-1.8$  &    $4.85$ &  $1.01\E{-2}$ & $ 503$ &  $6.78\E{-2}$ &  $2.71\E{-3}$ &  $9.98\E{-3}$ &       $0.139$  \\
\hline                                                                                                     
{\tt     M25E1}     &   $25$ &    $1$ &    $1.69$ &    $1.89$ & $-\infty$ &    $1.89$ &  $7.00\E{-2}$ & $1118$ &  $4.44\E{-2}$ &       $0.108$ &       $0.145$ &       $0.636$ \\
{\tt  M25E1mol.fix} &   $25$ &    $1$ &    $1.69$ &    $1.89$ & $-\infty$ &    $1.89$ &  $7.00\E{-2}$ & $1118$ &  $4.15\E{-2}$ &       $0.188$ &       $0.153$ &       $0.674$ \\
{\tt    M25E1F}     &   $25$ &    $1$ &    $1.69$ &    $5.4$  &   $-2.2$  &    $5.38$ &  $1.54\E{-3}$ & $ 767$ &  $4.19\E{-2}$ &    $<10^{-5}$ &    $<10^{-5}$ &  $4.83\E{-2}$ \\
\enddata
\tablecomments{
(1) ID of runs.
(2) Progenitor mass.
(3) Explosion energy.
(4) Initial mass cut.
(5) Outer boundary of mixing.
(6) Ejection fraction.
(7) Remnant mass.
(8) Ejected $^{56}$Ni mass.
(9) Time when we terminate the calculations in days.
(10-13) Mass of C, Si, and $\Forsterite$ grains and total dust mass.}
\end{deluxetable*}


\section{Methods} \label{sec:methods}

We use a one-dimensional spherically symmetric stellar and explosive nucleosynthesis model calculated by 
\citet{tominaga07}.
The electron fraction $Y_{\rm e}$ in the incomplete Si-burning region is set to be 0.4997 \citep{kobayashi06}.
The model includes reaction networks of 280 isotopes up to $^{79}$Br.
We then calculate the final abundance of isotopes, considering the decay of radioactive isotopes.
Summing up the isotopes with the same atomic number at a given radius, we obtain the radial distribution of elemental abundances as the initial condition of our molecule/dust formation calculations.

The density and temperature evolutions of SN ejecta are calculated in detail, at each mass coordinate in the expanding ejecta, using the method developed by \citet{Iwamoto00}.
Assuming a homologous expansion, 
the density $\rho (\Mr)$ at each mass coordinate $\Mr$ decreases as $\rho (\Mr) \propto t^{-3}$ as a function of time $t$.
We include adiabatic expansion cooling and $^{56}$Ni decay heating.
The model also solves radiation transfer, and
the gas loses and gains thermal energy through thermal radiation and photon absorption, respectively.

At each mass coordinate, our model solves the formation/dissociation of 8 diatomic molecular species: SiC, CO, SiS, SiO, O$_2$, S$_2$, CS, and SO.
We calculate the radiative association reaction rates with the empirical Arrhenius form.
Dissociation is induced by the collision with all gas-phase species whose kinetic energy exceeds the binding energy of the dissociating molecules.
We calculate the reaction cross-sections from the geometrical radii of impactors.

Our model simultaneously solves the nucleation and growth for 12 grain species: C, Si, SiC, Fe, FeS, FeO, $\Magnetite$, $\Alumina$, $\Enstatite$, $\Forsterite$, $\Silica$, and MgO, using the non-steady state classical nucleation theory (CNT) of \citetalias{Nozawa13} \citep[see also][for a review]{Schneider23}.\footnote{Although the UV bump in the galaxy JADES-GS-z6-0 is the sign of graphite, we consider carbonaceous grains as monoatomic, amorphous carbon grains in this work.}
In this model, we assume that grains form through the reactions presented in Table 2 of \citet{Nozawa03}.
For instance, alumina ($\Alumina$) is produced by the reaction
\begin{equation}
{\rm 2 Al (g) + 3 O (g) \to {\Alumina} (s)},
\label{eq:alumina}
\end{equation}
where (g) and (s) denote the gas and solid phase, respectively.
Through the reaction, in a mass coordinate with a sufficient abundance of Al and O, these atoms can directly condense into alumina grains when the supersaturation rate becomes below $\sim 1$ (see more in discussion).
Also, we here adopt so-called a ``key-species'' approach, where the reaction rate is limited by the minimum collision rate among reactant species (key species).
After the reaction, the abundance of other reactants decreases by the stoichiometric number times the decrement of the key species.

In this work, we use a hypernova model with a progenitor mass of $\MPopIII = 25~\Msun$ and explosion energy of $10$ B ($1~{\rm B} = 10^{51}~{\rm erg}$).
\citet{Ishigaki18} compared the elemental abundances of 202 EMP stars with the yield of Pop III SNe with different $\MPopIII$ and $\Esn$.
Then they found that the majority of the EMP stars (55\%) can be fitted with the Pop III hypernova model with $\MPopIII = 25~\Msun$ and $\Esn = 10$ B.
We also test the case with $\Esn = 1$ B (Section \ref{sec:lower_energy}).

Using the SN model, we run the following four calculations of dust synthesis as summarized in Table \ref{tab:models}.
In the first two runs, we use a mass cut for $\abM{^{56}Ni} = 0.07~\Msun$, by setting a remnant mass $\abM{rem} = 2.39~\Msun$.
The dotted curves in Fig. \ref{fig:rx_ini} (a) show the mass fraction of elements (after decays) as a function of mass coordinate in the SN ejecta.
Heavier elements, such as iron, are synthesized in the inner region.
The region below the plotted range ($M<2.39M_\odot$) are not ejected and falls into the remnant in the mass-cut model M25E10.
The blue line in Fig. \ref{fig:rx_ini} (b) shows the elemental abundances of the SN ejecta relative to C. 
With this mass-cut model, we run the following two calculations with different chemical processes to demonstrate the impact of molecule formation.
In the first case, C and Si are initially in the atomic form, and both molecule and dust formation are solved with our code ({\tt M25E10}).
In the second case, we assume a simplified model, where C and Si atoms are depleted into CO and SiO from the beginning, respectively, and the abundances of these molecules are fixed throughout the calculation ({\tt M25E10mol.fix}).

In the other two runs,
we use more realistic SN models
considering two mixing/fallback cases that correspond to the two major classifications of observed EMP stars:
one with solar-like (``normal'') elemental abundance ratios (C-normal stars) and the other with carbon enhancement of $\abFe{C} > 0.7$  \citep[C-enhanced metal-poor or CEMP stars;][]{Beers05, Chiaki17}.
Here we adopt a jet-like SN explosion model 
described with
three parameters, which are determined from the observations: the initial mass cut $\Mcut$, the fraction $\fej$ of gas ejected in the polar directions, and the mass $\Mmix$ that subjects to mixing \citep{Tominaga09}.
These parameters give the corresponding remnant mass $\Mrem = \Mcut + (1-\fej) (\Mmix - \Mcut)$.

In the former (called {\tt M25E10N}), we set a moderate $\Mmix = 3.1~\Msun$ and $\log \fej = -0.9$ to reproduce the elemental abundances of a C-normal star ${\rm SMSS~J093209.41-003435.8}$.
In the latter ({\tt M25E10F}), we set a large $\Mmix = 4.9$ and small $\log \fej = -1.8$ to reproduce those of a C-enhanced star ${\rm SMSS~J005953.98-594329.9}$ \citep{Ishigaki18}.
The dashed and solid curves in Fig. \ref{fig:rx_ini} (a) show the radial profile of elemental abundances for {\tt M25E10N} and {\tt M25E10F}, respectively, where
the inner part of ejecta ($\Mr < 2.92$ and $4.85~\Msun$) falls back on to the remnant.
The resultant yields are shown Fig. \ref{fig:rx_ini} (b) with green and red lines, respectively.
The latter type of SNe can eject Fe-deficient or relatively C-rich materials, 
and is used to reproduce the elemental abundances of C-enhanced metal-poor stars.
This SN type is called a `faint' SN due to its small $\abM{^{56}Ni}$. 
In these mixing/fallback models, we solve molecule formation reactions together with dust formation.

\begin{figure} 
\epsscale{1.17}
\plotone{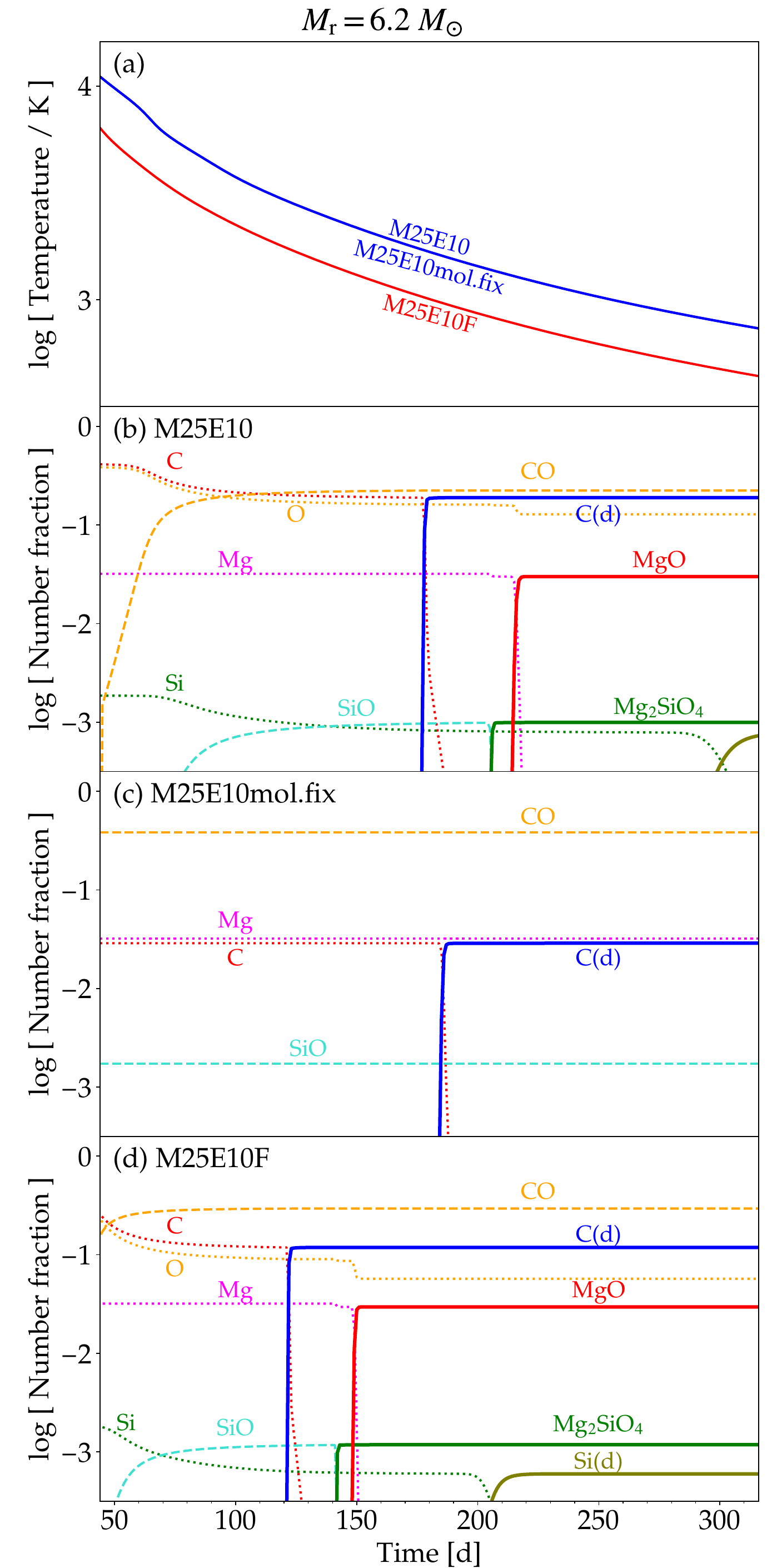}
\caption{Time evolution of (a) the gas temperature and (b--d) number abundance of chemical species at an enclosed mass of $M_{\rm r} = 6.2~\Msun$.
In the panel (a), the blue and red curves show the results for {\tt M25E10} and {\tt M25E10F}, respectively.
The curve for {\tt M25E10mol.fix} is overlapped with {\tt M25E10} because we use the same SN model.
The other panels indicate atomic (dotted), molecular (dashed), and grain (solid) species
for (b) {\tt M25E10}, (c) {\tt M25E10mol.fix}, and (d) {\tt M25E10F}.
We do not show the result for {\tt M25E10N} because it is almost the same as {\tt M25E10}.
\label{fig:ta}}
\end{figure}

In our code, we first calculate the temperature evolution in ejecta until 2000 d after explosion.
Using the results, we start the dust formation calculations when the temperature of at least one radius bin becomes below $10^4$ K and terminate the calculations when the temperature of all bins becomes below 300 K as shown in the 9th column of Table \ref{tab:models}.
Although radiative cooling due to newly forming molecules can affect the hydrodynamical evolution of ejecta \citep{Liljegren20}, we do not take it into account in this work to isolate the effect of molecule formation.

\vspace{1cm}
\begin{deluxetable*}{cccccccccccc}
\tabletypesize{\scriptsize}
\tablewidth{0pt} 
\tablecaption{Metal and grain production in {\tt M25E10} \label{tab:results}}
\tablehead{
\multicolumn{11}{c}{(a) Dust mass in units of $\Msun$} \\
\cline{1-11}
\colhead{  C} & 
\colhead{ Si} & 
\colhead{ Fe} & 
\colhead{FeS} & 
\colhead{$\Magnetite$} & 
\colhead{$\Alumina$  } & 
\colhead{$\Enstatite$} & 
\colhead{$\Forsterite$} & 
\colhead{$\Silica$} & 
\colhead{  MgO} & 
\colhead{Total}
} 
\startdata
         $8.45\E{-2}$ &              $0.196$ &         $2.05\E{-2}$ &         $9.73\E{-2}$ &         $2.14\E{-3}$ &         $1.69\E{-3}$ &         $3.81\E{-3}$ &              $0.260$ &              $0.169$ &         $9.17\E{-2}$ &              $0.926$ \\
\enddata
\end{deluxetable*}
\vspace{-1cm}
\begin{deluxetable*}{cccccccc}
\vspace{-1.2cm}
\tabletypesize{\scriptsize}
\tablewidth{0pt} 
\tablehead{
\multicolumn{8}{c}{(b) Fraction of elements depleted into grains} \\
\cline{1-8}
\colhead{C} & 
\colhead{O} & 
\colhead{Mg} & 
\colhead{Al} & 
\colhead{Si} & 
\colhead{S} & 
\colhead{Fe} & 
\colhead{Total}
} 
\startdata
              $0.304$ &              $0.103$ &              $0.939$ &              $1.000$ &              $0.738$ &              $0.153$ &              $1.000$ &              $0.234$ \\
\enddata
\end{deluxetable*}

\begin{figure*} 
\epsscale{1.17}
\plotone{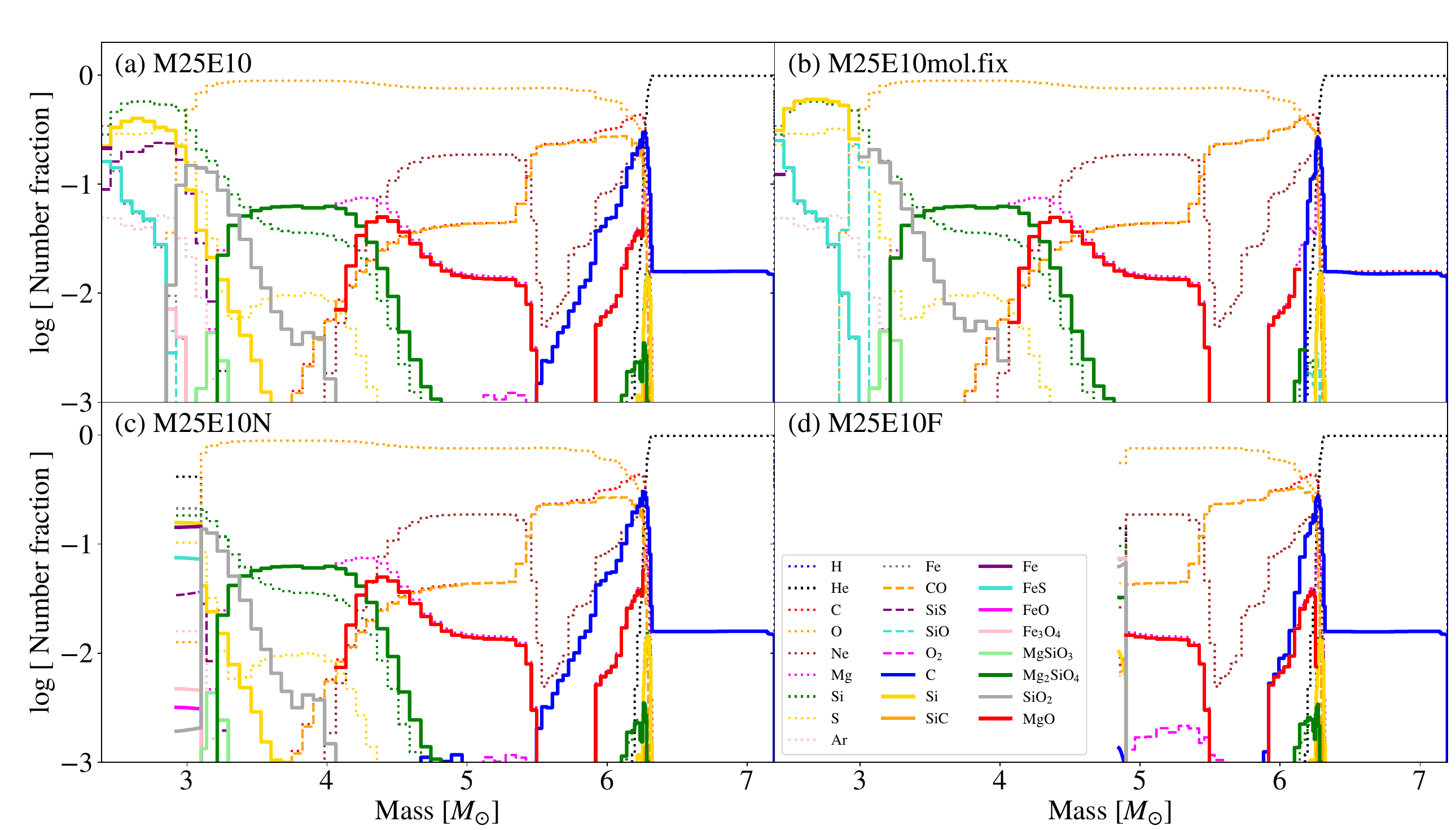}
\caption{Number abundance of chemical species as a function of enclosed mass for (a) {\tt M25E10}, (b) {\tt M25E10mol.fix}, (c) {\tt M25E10N}, and (d) {\tt M25E10F}.
The dotted curves indicate the initial abundances of elements, and the dashed and solid curves indicate the final abundances of molecules and grains, respecively.
\label{fig:rx}}
\end{figure*}

\section{Results} \label{sec:results}

We can successfully calculate molecule and dust formation in the ejecta fully self-consistently as Fig.\,\ref{fig:ta}.
Table\,\ref{tab:models} gives the summary of model parameters and results.
In this section, we in detail analyze the produced dust mass and its composition.

\subsection{Molecule and dust formation}
We first present the results in the model {\tt M25E10}.
Fig.\,\ref{fig:ta} shows the evolution of (a) temperature and (b) species abundances
at an enclosed mass of $\Mr = 6.2~\Msun$ (C layer).
At the time $t=80$ d, a part of C atoms (red dotted) are oxidized into CO molecules (orange dashed). 
At $t=177$ d where the gas temperature becomes below $\sim 1700$ K, which is comparable to the sublimation temperature, C grains (blue solid) start to form from remaining C atoms.\footnote{In Fig.\,\ref{fig:ta}, the blue solid curve shows the number fraction of C nuclei locked up into C grains.}
Fig.\,\ref{fig:rx} (a) shows the number fraction of chemical species as a function of $\Mr$. 
C grains (blue solid) form at a range of $\Mr = 5.5$--$6.3~\Msun$, and in total $8.45\E{-2}~\Msun$ C grains are ejected.

The other species form in different layers with different elemental abundances (Fig.\,\ref{fig:rx}).
Metallic Si (yellow solid) forms mainly in the Si-S-Fe layer ($M_{\rm r} = 2.5$--$3.0~\Msun$). 
A fraction of Si atoms that are not depleted into SiS molecules can condense into metallic Si grains.
In the O-Si-S layer ($\Mr = 3.0$--$3.2~\Msun$), $\Silica$ is the dominant grain species.
$\Forsterite$ (green solid) forms mainly in the O-Si-Mg and O-Mg-Si layers ($M_{\rm r} = 3.2$--$4.4~\Msun$). 

We show the mass of the major dust species in Table\,\ref{tab:results} (a).
The dominant species are metallic Si, $\Forsterite$, and $\Silica$.
In total $\sim 1~\Msun$ dust is produced in {\tt M25E10}.
Table \ref{tab:results} (b) shows the mass fraction of elements finally depleted into grains.
We find that $\sim 30$\% of C nuclei are locked up into C grains 
while almost 100\% of Mg, Al, and Fe are depleted into grains.
In total, 23.4\% of metals are locked up into grains.


\subsection{The importance of molecule formation}

In the simple chemistry model, {\tt M25E10mol.fix}, the total dust mass ($\Mdust = 0.985~\Msun$) is almost the same as {\tt M25E10mol}, but the masses of several species, such as C and Si, are significantly different.
The C grain mass is underestimated by $\sim 30$\%, compared with {\tt M25E10}.
We can clearly see that the C grain abundance is smaller in the inner region of the C layer ($\Mr = 5.5$--$6.2~\Msun$), comparing Figs. \ref{fig:rx} (a) and (b). 
To interpret this, we compare the evolution of species abundances for {\tt M25E10} and {\tt M25E10mol.fix} at $\Mr = 6.2~\Msun$ (Fig. \ref{fig:ta}b and c).
In this layer, the C abundance is only slightly larger than O, and thus only the small fraction of C atoms ($\sim 0.03$) are available to condense in {\tt M25E10mol.fix} (panel c). 
In {\tt M25E10} (panel b), C and O are initially in the atomic form, and CO molecules form at a timescale comparable to grain formation.
Therefore, the larger amount of C atoms ($\sim 0.2$) remain just before C grain formation.


Metallic Si mass is overestimated in {\tt M25E10mol.fix} by $\sim 40$\%, compared with {\tt M25E10}.
In {\tt M25E10mol.fix}, where we consider only SiO molecule formation,
all Si can remain in the atomic form because O is not synthesized in the Si-rich layer (Fig. \ref{fig:rx_ini}).
In {\tt M25E10}, since we include the formation of molecules other than SiO,
a part of Si has been depleted into SiS molecules before Si grain formation.



\subsection{Mixing/fallback}
In the mass-cut model with $\abM{^{56}Ni} = 0.07~\Msun$, the mass ratio of C grains to silicates is small ($\sim 0.325$) because the ejecta is overall O-rich.
In the extinction curve derived from the grain composition, the observed UV bump would not appear.
In mixing/fallback models, however, only silicate grain mass is expected to be suppressed.

In {\tt M25E10N}, mixing/fallback can reduce the total dust mass down to $0.706~\Msun$, compared with {\tt M25E10}, due to the fallback of inner materials (Table \ref{tab:models}).
However, the C grain mass is almost unchanged ($8.70\E{-2}~\Msun$) because the outer C-layer is not affected by 
mixing/fallback.
The Fe-Si-S and Si-S-Fe layers ($\Mr < 3.0~\Msun$) fall back (dashed curves in Fig. \ref{fig:rx_ini}), which suppresses the mass of Si and FeS grains by $\sim 50$\%.
However, silicate mass is still high ($0.259~\Msun$), and the ratio of C grains 
to silicates is only $\sim 0.335$.
In this case, the interstellar medium enriched by this SN still would not show the remarkable UV bump.

In {\tt M25E10F}, the inner region even within the O-Ne-Mg layer ($\Mr < 4.85~\Msun$) falls back.
Accordingly, the masses of Fe-, Si-, Mg-, Al-, and O-baring species are reduced by several orders of magnitude (Fig.\,\ref{fig:rx}d).
On the other hand, 
the mass of C grains produced is
$6.78\E{-2}~\Msun$, only $\sim 20$\% lower than in {\tt M25E10}.

The C grain production is not significantly changed because the C layer ($\Mr \sim 6~\Msun$) 
is not affected by mixing/fallback.
The slight decrease by $\sim 20$\% is due to the temperature evolution of the ejecta (Fig. \ref{fig:ta}d).
In this faint SN model, the mass of the main heat source, $^{56}$Ni, is smaller ($\abM{^{56}Ni} = 1.01\E{-2}~\Msun$) than the other models.
Since the gas temperature is lower than in {\tt M25E10} by $\sim 20$\%, CO molecule formation is slightly enhanced.
Then, available C atoms are reduced, and thus the mass of C grains is slightly suppressed.
The mass ratio of C grains to silicates is still high ($\sim 7$), that can explain the observed spectral feature of the galaxy JADES-GS-z6-0.



\subsection{Lower explosion energy case} \label{sec:lower_energy}
So far we have set the explosion energy to be $\Esn = 10$ B.
The properties of forming grains may be altered by the explosion energy.
We also test the case with a lower explosion energy, $\Esn = 1$ B, keeping the other parameters the same
as in Table 1. 
With $0.07~\Msun$ $^{56}$Ni, molecule formation self-consistently solved in {\tt M25E1} but not in {\tt M25E1mol.fix}.
Finally we consider a mixing/fallback model, {\tt M25E1F}, with $\Mmix = 5.4~\Msun$ and $\log \fej = -2.2$ that can reproduce the elemental abundance of a C-enhanced star, SDSS J014036.21+234458.1.

As summerized in Table \ref{tab:models}, the dust mass for {\tt M25E1} ($0.636~\Msun$) is smaller than in {\tt M25E10}.
This is because the synthesized metal mass is smaller for the lower explosion energy.
We find that the C grain mass is almost the same between {\tt M25E1} and {\tt M25E1mol.fix} unlike the $\Esn = 10$ B case.
This is because the grain formation timescale is longer relative to the CO molecule formation timescale than in {\tt M25E10}.
With the lower explosion energy, ejecta expands more slowly, and the gas temperature becomes below the sublimation temperature later.

For {\tt M25E1F}, the dust mass ($4.27\E{-2}~\Msun$) is smaller than in {\tt M25E1} ($\sim 0.6~\Msun$) because of the fallback. 
The Si and $\Forsterite$ grain masses are much smaller than in {\tt M25E1}, while the C grain mass ($4.19\E{-2}~\Msun$) is almost unchanged from {\tt M25E1} ($4.44\E{-2}~\Msun$). 
These tendencies are the same as in the $\Esn = 10$ B case, and thus we can conclude that both faint SNe and faint hypernovae can cause the remarkable UV bump.

\section{Discussion}

Although our dust synthesis code is unique with the self-consistent treatment of molecule formation, our results can be compared with other studies as follows.
\citet{Brooker22} studied dust formation with the same formulation as in our work, but without molecule formation and with different initial metallicities of $10^{-4}$--$10^{-1}~\Zsun$. 
Since the metals synthesized in the pre-SN phase are dominant over the initially existing metals, the initial metallicity should have only a minor effect on the dust mass.
They run simulations with a progenitor mass of $25~\Msun$ with $\sim$10~B and $
\sim$1~B, which are comparable to our models without molecule formation. 
With 10B, the C grain mass in our {\tt M25E10mol.fix} model ($5.77\E{-2}~\Msun$) is a factor of four larger than the result of their {\tt M25bE9.73} model ($1.39\E{-2}~\Msun$).
The silicate mass in our {\tt M25E10} model ($0.260~\Msun$) is very similar to their result ($0.256~\Msun$).
With 1 B, the C grain mass in our {\tt M25E1mol.fix} model ($4.15\E{-2}~\Msun$) is a factor of three larger than the result of their {\tt M25d3E0.92} model ($1.37\E{-2}~\Msun$).
The silicate mass obtained by our study ($0.153~\Msun$) is similar to their result ($0.208~\Msun$).
Therefore, the carbonaceous dust mass is larger in our models than in \citet{Brooker22}. This is  because the abundance of C atoms is smaller than in their model in the C-rich layer.

These studies of dust formation are based on the CNT-based approach as well as this work.
Another approach is to compute non-equilibrium chemistry related to the formation of molecules and dust clusters \citep{Sarangi13, Biscaro14}.
The grain mass with this approach is smaller than with our CNT approach by a factor of $\sim$10.
This may be because high-energy ions destroy the reactant molecules.
For example, in their model, $\Alumina$ grains form through the reaction including AlO molecules that are not included in our model (Eq.\ref{eq:alumina}).
The molecules tend to be quickly destroyed by Ne$^+$ ions in the gas.
It is required to carefully compare these approaches in further detail.

There are longer-term physical processes that 
can affect the dust mass produced by a SN.
First of all, if SN ejecta interact with the dense circumstellar medium (CSM), dust can form again in the interaction region at later time \citep[$\sim 10$ yr;][]{Shahbandeh25}.
In this work, dust formation is followed only until $\sim 1000$ days, and additional dust production in the CSM is not considered. 
Note that, however, our model can be applied to the CSM interaction once its physical condition is specified.
After that, reverse shocks propagating in ejecta are supposed to destroy a significant amount of dust produced by a SN $\sim\!10^4$ yr after the explosion, which can reduce the final dust mass released from the SN \citep{Bianchi07, Nozawa07}.

Secondly, dust mass could be enhanced by a clumpy structure of ejecta, although spherically symmetric ejecta structure is assumed in this paper. 
\citet{Sarangi15} found that the formation of large grains ($\gtrsim 0.1~\um$) is favored in clumpy ejecta, and the mass of several grain species, such as metallic silicon, can be enhanced.
It would be interesting to test grain formation with clumpy ejecta structure particularly in HN cases, which might increase the metallic silicon mass and make it dominant species.


Finally, in this paper, we focused on the mass fraction of carbonaceous grains produced by a Pop III SN and have not calculated their sizes. 
The size distribution is crucial to the cooling efficiency of star-forming clouds \citep{Chiaki19, Chiaki20} as well as the extinction curves of galaxies \citep{dubois24}, which we hope to predict in our future works.

\section{Conclusion}

A recent JWST observation have revealed the spectral feature (``UV bump'') of carbonaceous grains in the distant galaxy JADES-GS-z6-0 at $z=6.7$ \citep{Witstok23a}.
It is puzzling how the galaxy have acquired C grains in the very early universe.
In this paper, we present the first results of our state-of-the-art dust synthesis code to study dust formation in Pop III SN ejecta, including molecule formation and mixing/fallback effects, fully self-consistently.

First, we find that the inclusion of molecule formation reactions significantly affect grain formation.
It is oversimplification if 
all available C and Si are assumed to deplete into CO and SiO initially, as in previous works.
C grain mass is underestimated by $\sim 40$\%, compared to our fiducial model where both molecule and dust formation reactions are solved self-consistently.
This is because the timescales for CO molecular formation and C grain formation is comparable, and a part of C atoms can remain just before C grain formation.
Larger dust production per SN might help explaining the large dust mass observed with ALMA.
However, a large amount of silicates 
is produced at the same time, which is likely to hide the UV bump. 

Secondly, we find that the mixing/fallback of ejecta remarkably affects the dust composition.
We consider two models that reproduce the elemental abundances of descendant C-normal and C-enhanced metal-poor stars ({\tt M25E10N} and {\tt M25E10F}, respectively).
%
For the faint SN model, 
although the total dust amount is 85\% smaller, the C grain mass ($6.78\E{-2}~\Msun$) is only 20\% smaller than in the hypernova model, resulting the
carbonaceous to silicate grains ratio of $\sim 7$.
This may be a clue to explain 
the remarkable UV bump 
in a high-redshift galaxy \citep{Witstok23a}.
We conclude that faint SNe can be the C grain factory in the early universe.

This study is the very first step to uncover the cosmic dust formation history.
In this work, we showed the dust masses only for $\MPopIII = 25~\Msun$ and explosion energies $\Esn = 1$ and $10$ B.
For higher $\MPopIII$, the dust mass would be larger because the synthesized metal mass is larger \citep{Nozawa03}.
For higher $\Esn$, the more abundant O is synthesized.
Therefore, we can expect the mass of oxidized grains, such as silicates, silica, and alumina increases.
In future works, we will study dust formation in ejecta of various progenitors, ideally taking account of the multi-dimensional and clumpy effects on the additional formation and destruction in the CSM (\S 4).
From these studies, we will eventually build a comprehensive GCE model to self-consistently follow the cosmic evolution of dust and metals.

\section*{Acknowledgments}

\begin{acknowledgments}
We are grateful to T. Nozawa for his crucial contributions to this research. 
We thank M. Ishigaki, T. Hosokawa, R. Matsukoba, and K. Omukai for discussions.
CK acknowledges funding from the UK Science and Technology Facility Council through grant ST/R000905/1, ST/V000632/1, ST/Y001443/1. The work was also funded by a Leverhulme Trust Research Project Grant on ``Birth of Elements''. 
\software{{\sc matplotlib} \citep{Hunter07}
          }
\end{acknowledgments}

\section*{Data Availability}

Two SN models for 25$\Msun$ models with explosion energies of 1 and 10 B are available on Zenodo at DOI: \href{https://zenodo.org/records/19364861}{10.5281/zenodo.19364861}. 
Each file provides the masses of the major element nuclei, the mass fraction of elements depleted onto dust grains, and the masses of molecules and dust grains at the end of the model simulation.

\vspace{5mm}




\bibliography{main}{}

@ARTICLE{Watson15,
       author = {{Watson}, Darach and {Christensen}, Lise and {Knudsen}, Kirsten Kraiberg and {Richard}, Johan and {Gallazzi}, Anna and {Micha{\l}owski}, Micha{\l} Jerzy},
        title = "{A dusty, normal galaxy in the epoch of reionization}",
      journal = {\nat},
         year = 2015,
        month = mar,
       volume = {519},
       number = {7543},
        pages = {327-330},
          doi = {10.1038/nature14164},
archivePrefix = {arXiv},
       eprint = {1503.00002},
 primaryClass = {astro-ph.GA},
       adsurl = {https://ui.adsabs.harvard.edu/abs/2015Natur.519..327W}
}

@ARTICLE{Tamura19,
       author = {{Tamura}, Yoichi and {Mawatari}, Ken and {Hashimoto}, Takuya and {Inoue}, Akio K. and {Zackrisson}, Erik and {Christensen}, Lise and {Binggeli}, Christian and {Matsuda}, Yuichi and {Matsuo}, Hiroshi and {Takeuchi}, Tsutomu T. and {Asano}, Ryosuke S. and {Sunaga}, Kaho and {Shimizu}, Ikkoh and {Okamoto}, Takashi and {Yoshida}, Naoki and {Lee}, Minju M. and {Shibuya}, Takatoshi and {Taniguchi}, Yoshiaki and {Umehata}, Hideki and {Hatsukade}, Bunyo and {Kohno}, Kotaro and {Ota}, Kazuaki},
        title = "{Detection of the Far-infrared [O III] and Dust Emission in a Galaxy at Redshift 8.312: Early Metal Enrichment in the Heart of the Reionization Era}",
      journal = {\apj},
         year = 2019,
        month = mar,
       volume = {874},
       number = {1},
          eid = {27},
        pages = {27},
          doi = {10.3847/1538-4357/ab0374},
archivePrefix = {arXiv},
       eprint = {1806.04132},
 primaryClass = {astro-ph.GA},
       adsurl = {https://ui.adsabs.harvard.edu/abs/2019ApJ...874...27T}
}

@ARTICLE{Witstok22,
       author = {{Witstok}, Joris and {Smit}, Renske and {Maiolino}, Roberto and {Kumari}, Nimisha and {Aravena}, Manuel and {Boogaard}, Leindert and {Bouwens}, Rychard and {Carniani}, Stefano and {Hodge}, Jacqueline A. and {Jones}, Gareth C. and {Stefanon}, Mauro and {van der Werf}, Paul and {Schouws}, Sander},
        title = "{Dual constraints with ALMA: new [O III] 88-{\ensuremath{\mu}}m and dust-continuum observations reveal the ISM conditions of luminous LBGs at z   7}",
      journal = {\mnras},
         year = 2022,
        month = sep,
       volume = {515},
       number = {2},
        pages = {1751-1773},
          doi = {10.1093/mnras/stac1905},
archivePrefix = {arXiv},
       eprint = {2207.00022},
 primaryClass = {astro-ph.GA},
       adsurl = {https://ui.adsabs.harvard.edu/abs/2022MNRAS.515.1751W}
}

@ARTICLE{Beers05,
       author = {{Beers}, Timothy C. and {Christlieb}, Norbert},
        title = "{The Discovery and Analysis of Very Metal-Poor Stars in the Galaxy}",
      journal = {\araa},
         year = 2005,
        month = sep,
       volume = {43},
       number = {1},
        pages = {531-580},
          doi = {10.1146/annurev.astro.42.053102.134057},
       adsurl = {https://ui.adsabs.harvard.edu/abs/2005ARA&A..43..531B}
}

@ARTICLE{Asplund09,
       author = {{Asplund}, Martin and {Grevesse}, Nicolas and {Sauval}, A. Jacques and {Scott}, Pat},
        title = "{The Chemical Composition of the Sun}",
      journal = {\araa},
         year = 2009,
        month = sep,
       volume = {47},
       number = {1},
        pages = {481-522},
          doi = {10.1146/annurev.astro.46.060407.145222},
archivePrefix = {arXiv},
       eprint = {0909.0948},
 primaryClass = {astro-ph.SR},
       adsurl = {https://ui.adsabs.harvard.edu/abs/2009ARA&A..47..481A}
}

@ARTICLE{Umeda03,
       author = {{Umeda}, Hideyuki and {Nomoto}, Ken'ichi},
        title = "{First-generation black-hole-forming supernovae and the metal abundance pattern of a very iron-poor star}",
      journal = {\nat},
         year = 2003,
        month = apr,
       volume = {422},
       number = {6934},
        pages = {871-873},
          doi = {10.1038/nature01571},
archivePrefix = {arXiv},
       eprint = {astro-ph/0301315},
 primaryClass = {astro-ph},
       adsurl = {https://ui.adsabs.harvard.edu/abs/2003Natur.422..871U}
}

@ARTICLE{Iwamoto00,
       author = {{Iwamoto}, Koichi and {Nakamura}, Takayoshi and {Nomoto}, Ken'ichi and {Mazzali}, Paolo A. and {Danziger}, I. John and {Garnavich}, Peter and {Kirshner}, Robert and {Jha}, Saurabh and {Balam}, David and {Thorstensen}, John},
        title = "{The Peculiar Type IC Supernova 1997EF: Another Hypernova}",
      journal = {\apj},
         year = 2000,
        month = may,
       volume = {534},
       number = {2},
        pages = {660-669},
          doi = {10.1086/308761},
archivePrefix = {arXiv},
       eprint = {astro-ph/9807060},
 primaryClass = {astro-ph},
       adsurl = {https://ui.adsabs.harvard.edu/abs/2000ApJ...534..660I}
}

@ARTICLE{Tominaga07,
       author = {{Tominaga}, Nozomu and {Umeda}, Hideyuki and {Nomoto}, Ken'ichi},
        title = "{Supernova Nucleosynthesis in Population III 13-50 M$_{solar}$ Stars and Abundance Patterns of Extremely Metal-poor Stars}",
      journal = {\apj},
         year = 2007,
        month = may,
       volume = {660},
       number = {1},
        pages = {516-540},
          doi = {10.1086/513063},
archivePrefix = {arXiv},
       eprint = {astro-ph/0701381},
 primaryClass = {astro-ph},
       adsurl = {https://ui.adsabs.harvard.edu/abs/2007ApJ...660..516T}
}

@ARTICLE{Tominaga09,
       author = {{Tominaga}, Nozomu},
        title = "{Aspherical Properties of Hydrodynamics and Nucleosynthesis in Jet-Induced Supernovae}",
      journal = {\apj},
         year = 2009,
        month = jan,
       volume = {690},
       number = {1},
        pages = {526-536},
          doi = {10.1088/0004-637X/690/1/526},
archivePrefix = {arXiv},
       eprint = {0711.4815},
 primaryClass = {astro-ph},
       adsurl = {https://ui.adsabs.harvard.edu/abs/2009ApJ...690..526T}
}

@ARTICLE{Tominaga14,
       author = {{Tominaga}, Nozomu and {Iwamoto}, Nobuyuki and {Nomoto}, Ken'ichi},
        title = "{Abundance Profiling of Extremely Metal-poor Stars and Supernova Properties in the Early Universe}",
      journal = {\apj},
         year = 2014,
        month = apr,
       volume = {785},
       number = {2},
          eid = {98},
        pages = {98},
          doi = {10.1088/0004-637X/785/2/98},
archivePrefix = {arXiv},
       eprint = {1309.6734},
 primaryClass = {astro-ph.SR},
       adsurl = {https://ui.adsabs.harvard.edu/abs/2014ApJ...785...98T}
}

@ARTICLE{Ishigaki18,
       author = {{Ishigaki}, Miho N. and {Tominaga}, Nozomu and {Kobayashi}, Chiaki and {Nomoto}, Ken'ichi},
        title = "{The Initial Mass Function of the First Stars Inferred from Extremely Metal-poor Stars}",
      journal = {\apj},
         year = 2018,
        month = apr,
       volume = {857},
       number = {1},
          eid = {46},
        pages = {46},
          doi = {10.3847/1538-4357/aab3de},
archivePrefix = {arXiv},
       eprint = {1801.07763},
 primaryClass = {astro-ph.SR},
       adsurl = {https://ui.adsabs.harvard.edu/abs/2018ApJ...857...46I}
}

@ARTICLE{Choplin19,
       author = {{Choplin}, Arthur and {Tominaga}, Nozomu and {Ishigaki}, Miho N.},
        title = "{Inferring the velocity of early massive stars from the abundances of extremely metal-poor stars}",
      journal = {\aap},
         year = 2019,
        month = dec,
       volume = {632},
          eid = {A62},
        pages = {A62},
          doi = {10.1051/0004-6361/201936187},
archivePrefix = {arXiv},
       eprint = {1910.01366},
 primaryClass = {astro-ph.SR},
       adsurl = {https://ui.adsabs.harvard.edu/abs/2019A&A...632A..62C}
}

@ARTICLE{Nozawa03,
       author = {{Nozawa}, Takaya and {Kozasa}, Takashi and {Umeda}, Hideyuki and {Maeda}, Keiichi and {Nomoto}, Ken'ichi},
        title = "{Dust in the Early Universe: Dust Formation in the Ejecta of Population III Supernovae}",
      journal = {\apj},
         year = 2003,
        month = dec,
       volume = {598},
       number = {2},
        pages = {785-803},
          doi = {10.1086/379011},
archivePrefix = {arXiv},
       eprint = {astro-ph/0307108},
 primaryClass = {astro-ph},
       adsurl = {https://ui.adsabs.harvard.edu/abs/2003ApJ...598..785N}
}

@ARTICLE{Nozawa07,
       author = {{Nozawa}, Takaya and {Kozasa}, Takashi and {Habe}, Asao and {Dwek}, Eli and {Umeda}, Hideyuki and {Tominaga}, Nozomu and {Maeda}, Keiichi and {Nomoto}, Ken'ichi},
        title = "{Evolution of Dust in Primordial Supernova Remnants: Can Dust Grains Formed in the Ejecta Survive and Be Injected into the Early Interstellar Medium?}",
      journal = {\apj},
         year = 2007,
        month = sep,
       volume = {666},
       number = {2},
        pages = {955-966},
          doi = {10.1086/520621},
archivePrefix = {arXiv},
       eprint = {0706.0383},
 primaryClass = {astro-ph},
       adsurl = {https://ui.adsabs.harvard.edu/abs/2007ApJ...666..955N}
}

@ARTICLE{Nozawa13a,
       author = {{Nozawa}, Takaya and {Fukugita}, Masataka},
        title = "{Properties of Dust Grains Probed with Extinction Curves}",
      journal = {\apj},
         year = 2013,
        month = jun,
       volume = {770},
       number = {1},
          eid = {27},
        pages = {27},
          doi = {10.1088/0004-637X/770/1/27},
archivePrefix = {arXiv},
       eprint = {1301.4024},
 primaryClass = {astro-ph.GA},
       adsurl = {https://ui.adsabs.harvard.edu/abs/2013ApJ...770...27N}
}

@ARTICLE{Nozawa13,
       author = {{Nozawa}, Takaya and {Kozasa}, Takashi},
        title = "{Formulation of Non-steady-state Dust Formation Process in Astrophysical Environments}",
      journal = {\apj},
         year = 2013,
        month = oct,
       volume = {776},
       number = {1},
          eid = {24},
        pages = {24},
          doi = {10.1088/0004-637X/776/1/24},
archivePrefix = {arXiv},
       eprint = {1308.1873},
 primaryClass = {astro-ph.SR},
       adsurl = {https://ui.adsabs.harvard.edu/abs/2013ApJ...776...24N}
}

@ARTICLE{Todini01,
       author = {{Todini}, Paolo and {Ferrara}, Andrea},
        title = "{Dust formation in primordial Type II supernovae}",
      journal = {\mnras},
         year = 2001,
        month = aug,
       volume = {325},
       number = {2},
        pages = {726-736},
          doi = {10.1046/j.1365-8711.2001.04486.x},
archivePrefix = {arXiv},
       eprint = {astro-ph/0009176},
 primaryClass = {astro-ph},
       adsurl = {https://ui.adsabs.harvard.edu/abs/2001MNRAS.325..726T}
}

@ARTICLE{Schneider03,
       author = {{Schneider}, R. and {Ferrara}, A. and {Salvaterra}, R. and {Omukai}, K. and {Bromm}, V.},
        title = "{Low-mass relics of early star formation}",
      journal = {\nat},
         year = 2003,
        month = apr,
       volume = {422},
       number = {6934},
        pages = {869-871},
          doi = {10.1038/nature01579},
archivePrefix = {arXiv},
       eprint = {astro-ph/0304254},
 primaryClass = {astro-ph},
       adsurl = {https://ui.adsabs.harvard.edu/abs/2003Natur.422..869S}
}

@ARTICLE{Bianchi07,
       author = {{Bianchi}, Simone and {Schneider}, Raffaella},
        title = "{Dust formation and survival in supernova ejecta}",
      journal = {\mnras},
         year = 2007,
        month = jul,
       volume = {378},
       number = {3},
        pages = {973-982},
          doi = {10.1111/j.1365-2966.2007.11829.x},
archivePrefix = {arXiv},
       eprint = {0704.0586},
 primaryClass = {astro-ph},
       adsurl = {https://ui.adsabs.harvard.edu/abs/2007MNRAS.378..973B}
}

@ARTICLE{Marassi14,
       author = {{Marassi}, S. and {Chiaki}, G. and {Schneider}, R. and {Limongi}, M. and {Omukai}, K. and {Nozawa}, T. and {Chieffi}, A. and {Yoshida}, N.},
        title = "{The Origin of the Most Iron-poor Star}",
      journal = {\apj},
         year = 2014,
        month = oct,
       volume = {794},
       number = {2},
          eid = {100},
        pages = {100},
          doi = {10.1088/0004-637X/794/2/100},
archivePrefix = {arXiv},
       eprint = {1409.4424},
 primaryClass = {astro-ph.SR},
       adsurl = {https://ui.adsabs.harvard.edu/abs/2014ApJ...794..100M}
}

@ARTICLE{Marassi15,
       author = {{Marassi}, Stefania and {Schneider}, Raffaella and {Limongi}, Marco and {Chieffi}, Alessandro and {Bocchio}, Marco and {Bianchi}, Simone},
        title = "{The metal and dust yields of the first massive stars}",
      journal = {\mnras},
         year = 2015,
        month = dec,
       volume = {454},
       number = {4},
        pages = {4250-4266},
          doi = {10.1093/mnras/stv2267},
archivePrefix = {arXiv},
       eprint = {1509.08923},
 primaryClass = {astro-ph.GA},
       adsurl = {https://ui.adsabs.harvard.edu/abs/2015MNRAS.454.4250M}
}

@ARTICLE{Marassi19,
       author = {{Marassi}, S. and {Schneider}, R. and {Limongi}, M. and {Chieffi}, A. and {Graziani}, L. and {Bianchi}, S.},
        title = "{Supernova dust yields: the role of metallicity, rotation, and fallback}",
      journal = {\mnras},
         year = 2019,
        month = apr,
       volume = {484},
       number = {2},
        pages = {2587-2604},
          doi = {10.1093/mnras/sty3323},
archivePrefix = {arXiv},
       eprint = {1812.00009},
 primaryClass = {astro-ph.SR},
       adsurl = {https://ui.adsabs.harvard.edu/abs/2019MNRAS.484.2587M}
}

@ARTICLE{Witstok23a,
       author = {{Witstok}, Joris and {Shivaei}, Irene and {Smit}, Renske and {Maiolino}, Roberto and {Carniani}, Stefano and {Curtis-Lake}, Emma and {Ferruit}, Pierre and {Arribas}, Santiago and {Bunker}, Andrew J. and {Cameron}, Alex J. and {Charlot}, Stephane and {Chevallard}, Jacopo and {Curti}, Mirko and {de Graaff}, Anna and {D'Eugenio}, Francesco and {Giardino}, Giovanna and {Looser}, Tobias J. and {Rawle}, Tim and {Rodr{\'\i}guez del Pino}, Bruno and {Willott}, Chris and {Alberts}, Stacey and {Baker}, William M. and {Boyett}, Kristan and {Egami}, Eiichi and {Eisenstein}, Daniel J. and {Endsley}, Ryan and {Hainline}, Kevin N. and {Ji}, Zhiyuan and {Johnson}, Benjamin D. and {Kumari}, Nimisha and {Lyu}, Jianwei and {Nelson}, Erica and {Perna}, Michele and {Rieke}, Marcia and {Robertson}, Brant E. and {Sandles}, Lester and {Saxena}, Aayush and {Scholtz}, Jan and {Sun}, Fengwu and {Tacchella}, Sandro and {Williams}, Christina C. and {Willmer}, Christopher N.~A.},
        title = "{Carbonaceous dust grains seen in the first billion years of cosmic time}",
      journal = {\nat},
         year = 2023,
        month = sep,
       volume = {621},
       number = {7978},
        pages = {267-270},
          doi = {10.1038/s41586-023-06413-w},
archivePrefix = {arXiv},
       eprint = {2302.05468},
 primaryClass = {astro-ph.GA},
       adsurl = {https://ui.adsabs.harvard.edu/abs/2023Natur.621..267W}
}

@ARTICLE{Witstok23b,
       author = {{Witstok}, Joris and {Jones}, Gareth C. and {Maiolino}, Roberto and {Smit}, Renske and {Schneider}, Raffaella},
        title = "{An empirical study of dust properties at the earliest epochs}",
      journal = {\mnras},
         year = 2023,
        month = aug,
       volume = {523},
       number = {2},
        pages = {3119-3132},
          doi = {10.1093/mnras/stad1470},
archivePrefix = {arXiv},
       eprint = {2305.09714},
 primaryClass = {astro-ph.GA},
       adsurl = {https://ui.adsabs.harvard.edu/abs/2023MNRAS.523.3119W}
}

@ARTICLE{Schneider23,
       author = {{Schneider}, Raffaella and {Maiolino}, Roberto},
        title = "{The formation and cosmic evolution of dust in the early Universe. I. Dust sources}",
      journal = {arXiv e-prints},
         year = 2023,
        month = sep,
          eid = {arXiv:2310.00053},
        pages = {arXiv:2310.00053},
          doi = {10.48550/arXiv.2310.00053},
archivePrefix = {arXiv},
       eprint = {2310.00053},
 primaryClass = {astro-ph.GA},
       adsurl = {https://ui.adsabs.harvard.edu/abs/2023arXiv231000053S}
}

@ARTICLE{Chiaki17,
       author = {{Chiaki}, Gen and {Tominaga}, Nozomu and {Nozawa}, Takaya},
        title = "{Classification of extremely metal-poor stars: absent region in A(C)-[Fe/H] plane and the role of dust cooling}",
      journal = {\mnras},
         year = 2017,
        month = nov,
       volume = {472},
       number = {1},
        pages = {L115-L119},
          doi = {10.1093/mnrasl/slx163},
archivePrefix = {arXiv},
       eprint = {1710.04365},
 primaryClass = {astro-ph.GA},
       adsurl = {https://ui.adsabs.harvard.edu/abs/2017MNRAS.472L.115C}
}

@ARTICLE{Chiaki18,
       author = {{Chiaki}, Gen and {Susa}, Hajime and {Hirano}, Shingo},
        title = "{Metal-poor star formation triggered by the feedback effects from Pop III stars}",
      journal = {\mnras},
         year = 2018,
        month = apr,
       volume = {475},
       number = {4},
        pages = {4378-4395},
          doi = {10.1093/mnras/sty040},
archivePrefix = {arXiv},
       eprint = {1801.01583},
 primaryClass = {astro-ph.GA},
       adsurl = {https://ui.adsabs.harvard.edu/abs/2018MNRAS.475.4378C}
}

@ARTICLE{Chiaki19,
       author = {{Chiaki}, Gen and {Wise}, John H.},
        title = "{Seeding the second star: enrichment from population III, dust evolution, and cloud collapse}",
      journal = {\mnras},
         year = 2019,
        month = jan,
       volume = {482},
       number = {3},
        pages = {3933-3949},
          doi = {10.1093/mnras/sty2984},
archivePrefix = {arXiv},
       eprint = {1808.09515},
 primaryClass = {astro-ph.GA},
       adsurl = {https://ui.adsabs.harvard.edu/abs/2019MNRAS.482.3933C}
}

@ARTICLE{Chiaki20,
       author = {{Chiaki}, Gen and {Wise}, John H. and {Marassi}, Stefania and {Schneider}, Raffaella and {Limongi}, Marco and {Chieffi}, Alessandro},
        title = "{Seeding the second star - II. CEMP star formation enriched from faint supernovae}",
      journal = {\mnras},
         year = 2020,
        month = sep,
       volume = {497},
       number = {3},
        pages = {3149-3165},
          doi = {10.1093/mnras/staa2144},
archivePrefix = {arXiv},
       eprint = {2007.06657},
 primaryClass = {astro-ph.GA},
       adsurl = {https://ui.adsabs.harvard.edu/abs/2020MNRAS.497.3149C}
}

@ARTICLE{Biscaro14,
       author = {{Biscaro}, Chiara and {Cherchneff}, Isabelle},
        title = "{Molecules and dust in Cassiopeia A. I. Synthesis in the supernova phase and processing by the reverse shock in the clumpy remnant}",
      journal = {\aap},
         year = 2014,
        month = apr,
       volume = {564},
          eid = {A25},
        pages = {A25},
          doi = {10.1051/0004-6361/201322932},
archivePrefix = {arXiv},
       eprint = {1401.5594},
 primaryClass = {astro-ph.SR},
       adsurl = {https://ui.adsabs.harvard.edu/abs/2014A&A...564A..25B}
}

@ARTICLE{Sarangi15,
       author = {{Sarangi}, Arkaprabha and {Cherchneff}, Isabelle},
        title = "{Condensation of dust in the ejecta of Type II-P supernovae}",
      journal = {\aap},
         year = 2015,
        month = mar,
       volume = {575},
          eid = {A95},
        pages = {A95},
          doi = {10.1051/0004-6361/201424969},
archivePrefix = {arXiv},
       eprint = {1412.5522},
 primaryClass = {astro-ph.SR},
       adsurl = {https://ui.adsabs.harvard.edu/abs/2015A&A...575A..95S}
}

@ARTICLE{dayal22,
       author = {{Dayal}, P. and {Ferrara}, A. and {Sommovigo}, L. and {Bouwens}, R. and {Oesch}, P.~A. and {Smit}, R. and {Gonzalez}, V. and {Schouws}, S. and {Stefanon}, M. and {Kobayashi}, C. and {Bremer}, J. and {Algera}, H.~S.~B. and {Aravena}, M. and {Bowler}, R.~A.~A. and {da Cunha}, E. and {Fudamoto}, Y. and {Graziani}, L. and {Hodge}, J. and {Inami}, H. and {De Looze}, I. and {Pallottini}, A. and {Riechers}, D. and {Schneider}, R. and {Stark}, D. and {Endsley}, R.},
        title = "{The ALMA REBELS survey: the dust content of z   7 Lyman break galaxies}",
      journal = {\mnras},
         year = 2022,
        month = may,
       volume = {512},
       number = {1},
        pages = {989-1002},
          doi = {10.1093/mnras/stac537},
archivePrefix = {arXiv},
       eprint = {2202.11118},
 primaryClass = {astro-ph.GA},
       adsurl = {https://ui.adsabs.harvard.edu/abs/2022MNRAS.512..989D}
}

@INPROCEEDINGS{Umeda00,
       author = {{Umeda}, Hideyuki and {Nomoto}, Ken'ichi and {Nakamura}, Takayoshi},
        title = "{Evolution and Nucleosynthesis of Metal-Free Massive Stars}",
    booktitle = {The First Stars},
         year = 2000,
       editor = {{Weiss}, Achim and {Abel}, Tom G. and {Hill}, Vanessa},
        month = jan,
        pages = {150},
          doi = {10.1007/10719504_27},
archivePrefix = {arXiv},
       eprint = {astro-ph/9912248},
 primaryClass = {astro-ph},
       adsurl = {https://ui.adsabs.harvard.edu/abs/2000fist.conf..150U}
}

@ARTICLE{Bunker23,
       author = {{Bunker}, Andrew J. and {Saxena}, Aayush and {Cameron}, Alex J. and {Willott}, Chris J. and {Curtis-Lake}, Emma and {Jakobsen}, Peter and {Carniani}, Stefano and {Smit}, Renske and {Maiolino}, Roberto and {Witstok}, Joris and {Curti}, Mirko and {D'Eugenio}, Francesco and {Jones}, Gareth C. and {Ferruit}, Pierre and {Arribas}, Santiago and {Charlot}, Stephane and {Chevallard}, Jacopo and {Giardino}, Giovanna and {de Graaff}, Anna and {Looser}, Tobias J. and {L{\"u}tzgendorf}, Nora and {Maseda}, Michael V. and {Rawle}, Tim and {Rix}, Hans-Walter and {Del Pino}, Bruno Rodr{\'\i}guez and {Alberts}, Stacey and {Egami}, Eiichi and {Eisenstein}, Daniel J. and {Endsley}, Ryan and {Hainline}, Kevin and {Hausen}, Ryan and {Johnson}, Benjamin D. and {Rieke}, George and {Rieke}, Marcia and {Robertson}, Brant E. and {Shivaei}, Irene and {Stark}, Daniel P. and {Sun}, Fengwu and {Tacchella}, Sandro and {Tang}, Mengtao and {Williams}, Christina C. and {Willmer}, Christopher N.~A. and {Baker}, William M. and {Baum}, Stefi and {Bhatawdekar}, Rachana and {Bowler}, Rebecca and {Boyett}, Kristan and {Chen}, Zuyi and {Circosta}, Chiara and {Helton}, Jakob M. and {Ji}, Zhiyuan and {Kumari}, Nimisha and {Lyu}, Jianwei and {Nelson}, Erica and {Parlanti}, Eleonora and {Perna}, Michele and {Sandles}, Lester and {Scholtz}, Jan and {Suess}, Katherine A. and {Topping}, Michael W. and {{\"U}bler}, Hannah and {Wallace}, Imaan E.~B. and {Whitler}, Lily},
        title = "{JADES NIRSpec Spectroscopy of GN-z11: Lyman-{\ensuremath{\alpha}} emission and possible enhanced nitrogen abundance in a z = 10.60 luminous galaxy}",
      journal = {\aap},
         year = 2023,
        month = sep,
       volume = {677},
          eid = {A88},
        pages = {A88},
          doi = {10.1051/0004-6361/202346159},
archivePrefix = {arXiv},
       eprint = {2302.07256},
 primaryClass = {astro-ph.GA},
       adsurl = {https://ui.adsabs.harvard.edu/abs/2023A&A...677A..88B}
}

@ARTICLE{DEugenio23,
       author = {{D'Eugenio}, Francesco and {Maiolino}, Roberto and {Carniani}, Stefano and {Curtis-Lake}, Emma and {Witstok}, Joris and {Chevallard}, Jacopo and {Charlot}, Stephane and {Baker}, William M. and {Arribas}, Santiago and {Boyett}, Kristan and {Bunker}, Andrew J. and {Curti}, Mirko and {Eisenstein}, Daniel J. and {Hainline}, Kevin and {Ji}, Zhiyuan and {Johnson}, Benjamin D. and {Looser}, Tobias J. and {Nakajima}, Kimihiko and {Nelson}, Erica and {Rieke}, Marcia and {Robertson}, Brant and {Scholtz}, Jan and {Smit}, Renske and {Venturi}, Giacomo and {Tacchella}, Sandro and {Uebler}, Hannah and {Willmer}, Christopher N.~A. and {Willott}, Chris},
        title = "{JADES: Carbon enrichment 350 Myr after the Big Bang in a gas-rich galaxy}",
      journal = {arXiv e-prints},
         year = 2023,
        month = nov,
          eid = {arXiv:2311.09908},
        pages = {arXiv:2311.09908},
          doi = {10.48550/arXiv.2311.09908},
archivePrefix = {arXiv},
       eprint = {2311.09908},
 primaryClass = {astro-ph.GA},
       adsurl = {https://ui.adsabs.harvard.edu/abs/2023arXiv231109908D}
}

@ARTICLE{kobayashi06,
       author = {{Kobayashi}, Chiaki and {Umeda}, Hideyuki and {Nomoto}, Ken'ichi and {Tominaga}, Nozomu and {Ohkubo}, Takuya},
        title = "{Galactic Chemical Evolution: Carbon through Zinc}",
      journal = {\apj},
         year = 2006,
        month = dec,
       volume = {653},
       number = {2},
        pages = {1145-1171},
          doi = {10.1086/508914},
archivePrefix = {arXiv},
       eprint = {astro-ph/0608688},
 primaryClass = {astro-ph},
       adsurl = {https://ui.adsabs.harvard.edu/abs/2006ApJ...653.1145K}
}

@ARTICLE{kobayashi20,
       author = {{Kobayashi}, Chiaki and {Karakas}, Amanda I. and {Lugaro}, Maria},
        title = "{The Origin of Elements from Carbon to Uranium}",
      journal = {\apj},
         year = 2020,
        month = sep,
       volume = {900},
       number = {2},
          eid = {179},
        pages = {179},
          doi = {10.3847/1538-4357/abae65},
archivePrefix = {arXiv},
       eprint = {2008.04660},
 primaryClass = {astro-ph.GA},
       adsurl = {https://ui.adsabs.harvard.edu/abs/2020ApJ...900..179K}
}

@ARTICLE{Liljegren20,
       author = {{Liljegren}, S. and {Jerkstrand}, A. and {Grumer}, J.},
        title = "{Carbon monoxide formation and cooling in supernovae}",
      journal = {\aap},
         year = 2020,
        month = oct,
       volume = {642},
          eid = {A135},
        pages = {A135},
          doi = {10.1051/0004-6361/202038116},
archivePrefix = {arXiv},
       eprint = {2008.03160},
 primaryClass = {astro-ph.SR},
       adsurl = {https://ui.adsabs.harvard.edu/abs/2020A&A...642A.135L}
}

@Article{Hunter07,
  Author    = {Hunter, J. D.},
  Title     = {Matplotlib: A 2D graphics environment},
  Journal   = {Computing in Science \& Engineering},
  Volume    = {9},
  Number    = {3},
  Pages     = {90--95},
  publisher = {IEEE COMPUTER SOC},
  doi       = {10.1109/MCSE.2007.55},
  year      = 2007
}

@ARTICLE{Carniani24,
       author = {{Carniani}, Stefano and {Hainline}, Kevin and {D'Eugenio}, Francesco and {Eisenstein}, Daniel J. and {Jakobsen}, Peter and {Witstok}, Joris and {Johnson}, Benjamin D. and {Chevallard}, Jacopo and {Maiolino}, Roberto and {Helton}, Jakob M. and {Willott}, Chris and {Robertson}, Brant and {Alberts}, Stacey and {Arribas}, Santiago and {Baker}, William M. and {Bhatawdekar}, Rachana and {Boyett}, Kristan and {Bunker}, Andrew J. and {Cameron}, Alex J. and {Cargile}, Phillip A. and {Charlot}, St{\'e}phane and {Curti}, Mirko and {Curtis-Lake}, Emma and {Egami}, Eiichi and {Giardino}, Giovanna and {Isaak}, Kate and {Ji}, Zhiyuan and {Jones}, Gareth C. and {Kumari}, Nimisha and {Maseda}, Michael V. and {Parlanti}, Eleonora and {P{\'e}rez-Gonz{\'a}lez}, Pablo G. and {Rawle}, Tim and {Rieke}, George and {Rieke}, Marcia and {Del Pino}, Bruno Rodr{\'\i}guez and {Saxena}, Aayush and {Scholtz}, Jan and {Smit}, Renske and {Sun}, Fengwu and {Tacchella}, Sandro and {{\"U}bler}, Hannah and {Venturi}, Giacomo and {Williams}, Christina C. and {Willmer}, Christopher N.~A.},
        title = "{Spectroscopic confirmation of two luminous galaxies at a redshift of 14}",
      journal = {\nat},
         year = 2024,
        month = sep,
       volume = {633},
       number = {8029},
        pages = {318-322},
          doi = {10.1038/s41586-024-07860-9},
archivePrefix = {arXiv},
       eprint = {2405.18485},
 primaryClass = {astro-ph.GA},
       adsurl = {https://ui.adsabs.harvard.edu/abs/2024Natur.633..318C}
}

@ARTICLE{Brooker22,
       author = {{Brooker}, Ezra S. and {Stangl}, Sarah M. and {Mauney}, Christopher M. and {Fryer}, C.~L.},
        title = "{Dependence of Dust Formation on the Supernova Explosion}",
      journal = {\apj},
         year = 2022,
        month = jun,
       volume = {931},
       number = {2},
          eid = {85},
        pages = {85},
          doi = {10.3847/1538-4357/ac57c3},
archivePrefix = {arXiv},
       eprint = {2103.12781},
 primaryClass = {astro-ph.HE},
       adsurl = {https://ui.adsabs.harvard.edu/abs/2022ApJ...931...85B}
}

@ARTICLE{Shahbandeh25,
       author = {{Shahbandeh}, Melissa and {Fox}, Ori D. and {Temim}, Tea and {Dwek}, Eli and {Sarangi}, Arkaprabha and {Smith}, Nathan and {Dessart}, Luc and {Nickson}, Bryony and {Engesser}, Michael and {Filippenko}, Alexei V. and {Brink}, Thomas G. and {Zheng}, WeiKang and {Szalai}, Tam{\'a}s and {Johansson}, Joel and {Rest}, Armin and {Van Dyk}, Schuyler D. and {Andrews}, Jennifer and {Ashall}, Chris and {Clayton}, Geoffrey C. and {De Looze}, Ilse and {DerKacy}, James M. and {Dulude}, Michael and {Foley}, Ryan J. and {Gezari}, Suvi and {Gomez}, Sebastian and {Gonzaga}, Shireen and {Indukuri}, Siva and {Jencson}, Jacob and {Kasliwal}, Mansi and {Lane}, Zachary G. and {Lau}, Ryan and {Law}, David and {Marston}, Anthony and {Milisavljevic}, Dan and {O'Steen}, Richard and {Pierel}, Justin and {Siebert}, Matthew and {Skrutskie}, Michael and {Strolger}, Lou and {Tinyanont}, Samaporn and {Wang}, Qinan and {Williams}, Brian and {Xiao}, Lin and {Yang}, Yi and {Zs{\'\i}ros}, Szanna},
        title = "{JWST/MIRI Observations of Newly Formed Dust in the Cold, Dense Shell of the Type IIn SN 2005ip}",
      journal = {\apj},
         year = 2025,
        month = jun,
       volume = {985},
       number = {2},
          eid = {262},
        pages = {262},
          doi = {10.3847/1538-4357/adce77},
archivePrefix = {arXiv},
       eprint = {2410.09142},
 primaryClass = {astro-ph.HE},
       adsurl = {https://ui.adsabs.harvard.edu/abs/2025ApJ...985..262S}
}

@ARTICLE{Sarangi13,
       author = {{Sarangi}, Arkaprabha and {Cherchneff}, Isabelle},
        title = "{The Chemically Controlled Synthesis of Dust in Type II-P Supernovae}",
      journal = {\apj},
         year = 2013,
        month = oct,
       volume = {776},
       number = {2},
          eid = {107},
        pages = {107},
          doi = {10.1088/0004-637X/776/2/107},
archivePrefix = {arXiv},
       eprint = {1309.5887},
 primaryClass = {astro-ph.SR},
       adsurl = {https://ui.adsabs.harvard.edu/abs/2013ApJ...776..107S}
}

@ARTICLE{dubois24,
       author = {{Dubois}, Yohan and {Rodr{\'\i}guez Montero}, Francisco and {Guerra}, Corentin and {Trebitsch}, Maxime and {Han}, San and {Beckmann}, Ricarda and {Yi}, Sukyoung K. and {Lewis}, Joseph and {Jang}, J.~K.},
        title = "{Galaxies with grains: unraveling dust evolution and extinction curves with hydrodynamical simulations}",
      journal = {\aap},
         year = 2024,
        month = jul,
       volume = {687},
          eid = {A240},
        pages = {A240},
          doi = {10.1051/0004-6361/202449784},
archivePrefix = {arXiv},
       eprint = {2402.18515},
 primaryClass = {astro-ph.GA},
       adsurl = {https://ui.adsabs.harvard.edu/abs/2024A&A...687A.240D}
}
\bibliographystyle{aasjournal}

\end{document}